\documentclass[12pt]{article} %[twocolumn]
\usepackage[top=29mm, bottom=40mm, left=25mm, right=25mm]{geometry}
\usepackage{cite} %for hyphened citations

\usepackage{graphicx}
\usepackage{amsmath, amssymb} 				% Needed for align environment
\usepackage{empheq}
\usepackage{fancybox}
\usepackage{mathrsfs} %for \mathscr
\usepackage{textcomp} % for \textinterrobang
\usepackage{mathtools}

\usepackage{epstopdf}

\usepackage{latexsym}
\usepackage{marvosym}
\usepackage{bm}	%This is for bold greek letter and such; command is \bm

\usepackage[mathcal]{euscript}
\usepackage{slashed}
\usepackage{tikz}
	\usetikzlibrary{decorations.markings}
	\usetikzlibrary{arrows}
\usepackage{indentfirst}

\numberwithin{equation}{section}

\usepackage{hyperref}
\hypersetup{
	unicode,	% use with \texorpdfstring
	colorlinks,
	citecolor=cyan,% filecolor=black,% 
	linkcolor=cyan, % urlcolor=black,% pdftex
	bookmarksopen=true,
	bookmarksopenlevel=\maxdimen,
	}

\usepackage[all]{xy}
\usepackage{tikz}
	\usetikzlibrary{decorations.markings}
	\usetikzlibrary{arrows}

\usepackage{amsthm} %Enable theorem environments
\theoremstyle{definition}

\theoremstyle{plain}

\def\K3{\mathrm K3}

\def\gl#1#2{$\mathrm{GL}(#1; {\bf #2})$}
\def\sl#1#2{$\mathrm{SL}(#1; {\bf #2})$}
\def\sp#1#2{$\mathrm{Sp}(#1; {\bf #2})$}
\def\spin#1#2{$\mathrm{Spin}(#1, #2)$}
\def\U#1{U({#1})}

\def\double #1{#1{\hbox{\kern-2pt $#1$}}}
\def\dd{\hbox{\,\Large$\triangleright$}}

\def\pp{{\mathchoice
          {%displaystyle
              \kern 1pt%
              \raise 1pt
              \vbox{\hrule width5pt height0.4pt depth0pt
                    \kern -2pt
                    \hbox{\kern 2.3pt
                          \vrule width0.4pt height6pt depth0pt
                          }
                    \kern -2pt
                    \hrule width5pt height0.4pt depth0pt}%
                    \kern 1pt
           }
            {%textstyle
              \kern 1pt%
              \raise 1pt
              \vbox{\hrule width4.3pt height0.4pt depth0pt
                    \kern -1.8pt
                    \hbox{\kern 1.95pt
                          \vrule width0.4pt height5.4pt depth0pt
                          }
                    \kern -1.8pt
                    \hrule width4.3pt height0.4pt depth0pt}%
                    \kern 1pt
            }
            {%scriptstyle
              \kern 0.5pt%
              \raise 1pt
              \vbox{\hrule width4.0pt height0.3pt depth0pt
                    \kern -1.9pt  %[e]=0.15pt
                    \hbox{\kern 1.85pt
                          \vrule width0.3pt height5.7pt depth0pt
                          }
                    \kern -1.9pt
                    \hrule width4.0pt height0.3pt depth0pt}%
                    \kern 0.5pt
            }
            {%scriptscriptstyle
              \kern 0.5pt%
              \raise 1pt
              \vbox{\hrule width3.6pt height0.3pt depth0pt
                    \kern -1.5pt
                    \hbox{\kern 1.65pt
                          \vrule width0.3pt height4.5pt depth0pt
                          }
                    \kern -1.5pt
                    \hrule width3.6pt height0.3pt depth0pt}%
                    \kern 0.5pt%}
            }
        }}

\def\mm{{\mathchoice
                       {%displaystyle
                             \kern 1pt
               \raise 1pt    \vbox{\hrule width5pt height0.4pt depth0pt
                                  \kern 2pt
                                  \hrule width5pt height0.4pt depth0pt}
                             \kern 1pt}
                       {%textstyle
                            \kern 1pt
               \raise 1pt \vbox{\hrule width4.3pt height0.4pt depth0pt
                                  \kern 1.8pt
                                  \hrule width4.3pt height0.4pt depth0pt}
                             \kern 1pt}
                       {%scriptstyle
                            \kern 0.5pt
               \raise 1pt
                            \vbox{\hrule width4.0pt height0.3pt depth0pt
                                  \kern 1.9pt
                                  \hrule width4.0pt height0.3pt depth0pt}
                            \kern 1pt}
                       {%scriptscriptstyle
                           \kern 0.5pt
             \raise 1pt  \vbox{\hrule width3.6pt height0.3pt depth0pt
                                  \kern 1.5pt
                                  \hrule width3.6pt height0.3pt depth0pt}
                           \kern 0.5pt}
                       }}

\def\ad{{\kern0.5pt
                   \alpha \kern-5.05pt
\raise5.8pt\hbox{$\textstyle.$}\kern 0.5pt}}

\def\bd{{\kern0.5pt
                   \beta \kern-5.05pt \raise5.8pt\hbox{$\textstyle.$}\kern 0.5pt}}

\def\qd{{\kern0.5pt
                   q \kern-5.05pt \raise5.8pt\hbox{$\textstyle.$}\kern 0.5pt}}
\def\Dot#1{{\kern0.5pt
     {#1} \kern-5.05pt \raise5.8pt\hbox{$\textstyle.$}\kern 0.5pt}}

\def\on#1#2{{\buildrel{\mkern2.5mu#1\mkern-2.5mu}\over{#2}}}
\def\dt#1{\on{\hbox{\bf .}}{#1}}                % (big) dot over: seeÀ   below  
\def\under#1#2{\mathop{\null#2}\limits_{#1}}	% accent under
\mathcode`\*="702A                  % * now always complex conjugate
\def\half{{\textstyle{1\over{\raise.1ex\hbox{$\scriptstyle{2}$}}}}}
\def\slap#1#2{\setbox0=\hbox{$#1{#2}$}#2\kern-\wd0{\hbox to\wd0{\hfil$#1{/}$\hfil}}}
\def\sla#1{\mathpalette\slap{#1}}		%slash: see ÖÊÊÊbelow

\catcode128=13 \def €{\"A}                 % Option-u A
\catcode129=13 \def {\AA}                 % Option-A
\catcode130=13 \def '{\c}           	   % Option-C (cedilla)
\catcode131=13 \def ƒ{\'E}                   % Option-e E
\catcode132=13 \def "{\~N}                   % Option-n N
\catcode133=13 \def …{\"O}                 % Option-u O
\catcode134=13 \def †{\"U}                  % Option-u U
\catcode135=13 \def ‡{\'a}                  % Option-e a
\catcode136=13 \def ˆ{\`a}                   % Option-`  a
\catcode137=13 \def ‰{\^a}                 % Option-i a
\catcode138=13 \def Š{\"a}                 % Option-u a
\catcode139=13 \def ‹{\~a}                   % Option-n a
\catcode140=13 \def Œ{\alpha}            % Option-a
\catcode141=13 \def {\chi}                % Option-c
\catcode142=13 \def Ž{\'e}                   % Option-e e
\catcode143=13 \def {\`e}                    % Option-`  e
\catcode144=13 \def {\^e}                  % Option-i e
\catcode145=13 \def '{\"e}                % Option-u e
\catcode146=13 \def '{\'\i}                 % Option-e i
\catcode147=13 \def "{\`\i}                  % Option-`  i
\catcode148=13 \def "{\^\i}                % Option-i i
\catcode149=13 \def •{\"\i}                % Option-u i
\catcode150=13 \def –{\~n}                  % Option-n n
\catcode151=13 \def —{\'o}                 % Option-e o
\catcode152=13 \def ˜{\`o}                  % Option-`  o
\catcode153=13 \def ™{\^o}                % Option-i o
\catcode154=13 \def š{\"o}                 % Option-u o
\catcode155=13 \def ›{\~o}                  % Option-n o
\catcode156=13 \def œ{\'u}                  % Option-e u
\catcode157=13 \def {\`u}                  % Option-`  u
\catcode158=13 \def ž{\^u}                % Option-i u
\catcode159=13 \def Ÿ{\"u}                % Option-u u
\catcode160=13 \def  {\tau}               % Option-t
\catcode162=13 \def ¢{\oplus}           % Option-4
\catcode163=13 \def £{\relax\ifmmode\expandafter\to\else\expandafter\itemize\fi} % Option-3
\catcode164=13 \def ¤{\subset}	  % Option-6
\catcode165=13 \def ¥{\infty}           % Option-8
\catcode166=13 \def ¦{\mp}                % Option-7
\catcode167=13 \def §{\sigma}           % Option-s
\catcode168=13 \def ¨{\rho}               % Option-r
\catcode169=13 \def ©{\gamma}         % Option-g
\catcode170=13 \def ª{\leftrightarrow} % Option-2 ; Option-E (acute) :
\catcode171=13 \def «{\relax\ifmmode\expandafter\acute\else\expandafter\'\fi}
\catcode172=13 \def ¬{\relax\ifmmode\expandafter\ddt\else\expandafter\"\fi}
\catcode173=13 \def ­{\equiv}            % Option-= ; ^ Option-U (umlaudt)
\catcode174=13 \def ®{{{}={}}}          % Option-"
\catcode175=13 \def ¯{\Omega}          % Option-O
\catcode176=13 \def °{\otimes}          % Option-5
\catcode177=13 \def ±{\ne}                 % Option-+
\catcode178=13 \def ²{\le}                   % Option-,
\catcode179=13 \def ³{\ge}                  % Option-.
\catcode180=13 \def ´{\upsilon}          % Option-y
\catcode181=13 \def µ{\mu}                % Option-m
\catcode182=13 \def ¶{\delta}             % Option-d
\catcode183=13 \def ·{\epsilon}          % Option-w
\catcode184=13 \def ¸{\Pi}                  % Option-P
\catcode185=13 \def ¹{\pi}                  % Option-p
\catcode186=13 \def º{\beta}               % Option-b
\catcode187=13 \def »{\partial}           % Option-9
\catcode188=13 \def ¼{\nobreak\ }       % Option-0
\catcode189=13 \def ½{\zeta}               % Option-z
\catcode190=13 \def ¾{\sim}                 % Option-'
\catcode191=13 \def ¿{\omega}           % Option-o
\catcode192=13 \def À{\dt}                     % Option-?
\catcode193=13 \def Á{\gets}                % Option-1
\catcode194=13 \def Â{\lambda}           % Option-l
\catcode195=13 \def Ã{\nu}                   % Option-v
\catcode196=13 \def Ä{\phi}                  % Option-f
\catcode197=13 \def Å{\xi}                     % Option-x
\catcode198=13 \def Æ{\psi}                  % Option-j
\catcode199=13 \def Ç{\int}                    % Option-\
\catcode200=13 \def È{\oint}                 % Option-|
\catcode201=13 \def É{\relax\ifmmode\expandafter\cdot\else\vol\fi}    % Option-;
\catcode202=13 \def Ê{\relax\ifmmode\expandafter\,\else\thinspace\fi}
\catcode203=13 \def Ë{\`A}                      % Option-`  A ; ^ Option-space
\catcode204=13 \def Ì{\~A}                      % Option-n A
\catcode205=13 \def Í{\~O}                      % Option-n O
\catcode206=13 \def Î{\Theta}              % Option-Q
\catcode207=13 \def Ï{\theta}               % Option-q; Option-- :
\catcode208=13 \def Ð{\relax\ifmmode\expandafter\bar\else\expandafter\=\fi}
\catcode209=13 \def Ñ{\overline}             % Option-_
\catcode210=13 \def Ò{\langle}               % Option-[
\catcode211=13 \def Ó{\relax\ifmmode\expandafter\{\else\ital\fi}      % Option-{
\catcode212=13 \def Ô{\rangle}               % Option-]
\catcode213=13 \def Õ{\}}                        % Option-}
\catcode214=13 \def Ö{\sla}                      % Option-/; Option-V :
\catcode215=13 \def ×{\relax\ifmmode\expandafter\check\else\expandafter\v\fi}
\catcode216=13 \def Ø{\"y}                     % Option-u y
\catcode217=13 \def Ù{\"Y}  		    % Option-u Y
\catcode218=13 \def Ú{\Leftarrow}       % Option-!
\catcode219=13 \def Û{\Leftrightarrow}       % Option-@ ; Option-# :
\catcode220=13 \def Ü{\relax\ifmmode\expandafter\Rightarrow\else\sect\fi}
\catcode221=13 \def Ý{\sum}                  % Option-$
\catcode222=13 \def Þ{\prod}                 % Option-%
\catcode223=13 \def ß{\widehat}              % Option-^
\catcode224=13 \def à{\pm}                     % Option-&
\catcode225=13 \def á{\nabla}                % Option-(
\catcode226=13 \def â{\quad}                 % Option-)
\catcode227=13 \def ã{\in}               	% Option-W
\catcode228=13 \def ä{\star}      	      % Option-R
\catcode229=13 \def å{\sqrt}                   % Option-M
\catcode230=13 \def æ{\^E}			% Option-i E
\catcode231=13 \def ç{\Upsilon}              % Option-Y
\catcode232=13 \def è{\"E}    	   	 % Option-u E
\catcode233=13 \def é{\`E}               	  % Option-`  E
\catcode234=13 \def ê{\Sigma}                % Option-S
\catcode235=13 \def ë{\Delta}                 % Option-D
\catcode236=13 \def ì{\Phi}                     % Option-F
\catcode237=13 \def í{\`I}        		   % Option-`  I
\catcode238=13 \def î{\iota}        	     % Option-H
\catcode239=13 \def ï{\Psi}                     % Option-J
\catcode240=13 \def ð{\times}                  % Option-K
\catcode241=13 \def ñ{\Lambda}             % Option-L
\catcode242=13 \def ò{\cdots}                % Option-:
\catcode243=13 \def ó{\^U}			% Option-i U
\catcode244=13 \def ô{\`U}    	              % Option-`  U
\catcode245=13 \def õ{\bo}                       % Option-B ; Option-I :
\catcode246=13 \def ö{\relax\ifmmode\expandafter\hat\else\expandafter\^\fi}
\catcode247=13 \def÷{\relax\ifmmode\expandafter\tilde\else\expandafter\~\fi}
\catcode248=13 \def ø{\ll}                         % Option-< ; ^ Option-N
\catcode249=13 \def ù{\gg}                       % Option-> 
\catcode250=13 \def ú{\eta}                      % Option-h 
\catcode251=13 \def û{\kappa}                  % Option-k 
\catcode252=13 \def ü{\half}     		 % Option-Z 
\catcode253=13 \def ý{\Gamma} 		% Option-G 
\catcode254=13 \def þ{\Xi}   			% Option-X ; Option-T : 
\catcode255=13 \def ÿ{\relax\ifmmode\expandafter{}^{\dagger}{}\else\dag\fi}

\def\S{{\cal S}}
\def\T{{\cal T}}
\def\U{{\cal U}}
\def\V{{\cal V}}

\title{
\color{violet} %{green}
{\bf F-theory with Worldvolume Sectioning}\\
}
\author{William D. Linch \textsc{iii}$\,{}^\text{\Pisces}$ and Warren Siegel$\,{}^\text{\Scorpio}$}
\date{}		% Activate to display a given date or no date

\begin{document}
\maketitle

%email addresses
\vspace*{-65mm}
\begin{flushright}
{%\color{red}
UMDEPP-015-006\\
YITP-SB-15-6
}
\end{flushright}
\vspace*{+40mm}

\begin{center}
%\vskip 0.2in
{\em
${}^{\mbox{\footnotesize\Pisces}}$
Center for String and Particle Theory,
Department of Physics,\\
University of Maryland at College Park,
College Park, MD 20742-4111\\
~\\
${}^{\mbox{\footnotesize\Scorpio}}$
C. N. Yang Institute for Theoretical Physics\\
State University of New York, Stony Brook, NY 11794-3840
}%\\
\end{center} 

\vspace{10pt}

\begin{abstract}
We describe the worldvolume for the bosonic sector of the lower-dimensional F-theory that embeds 5D, N=1 M-theory and the 4D type II superstring.
This theory is a complexification of the fundamental 5-brane theory that embeds the 4D, N=1 M-theory of the 3D type II string in a sense that we make explicit at the level of the Lagrangian and Hamiltonian formulations.
We find three types of section condition: in spacetime, on the worldvolume, and one tying them together.
The 5-brane theory is recovered from the new theory by a double dimensional reduction.
\end{abstract}

%email addresses
\vspace*{.5cm}
\begin{flushleft}
~\\
{${}^{\mbox{\footnotesize\Pisces}}$ \href{mailto:wdlinch3@gmail.com}{wdlinch3@gmail.com}}\\
%\makebox[0pt][t]
{$^{\text{\Scorpio}}$ \href{mailto:siegel@insti.physics.sunysb.edu}{siegel@insti.physics.sunysb.edu}}
%\makebox[0pt][t]
\end{flushleft}

\setcounter{page}0
\thispagestyle{empty}

\newpage

%\setcounter{page}0
%\thispagestyle{empty}
%\tableofcontents
%\newpage

%%%%%%%%%%%%%%%%%%%%%%%%%%%%%%%%%%%%%%%%%%%%%%%%%%%%%%%%%%%%%%%%
%%%%%%%%%%%%%%%%%%%%%%%%%%%%%%%%%%%%%%%%%%%%%%%%%%%%%%%%%%%%%%%%
\section{Introduction}

Previously \cite{Polacek:2014cva,Linch:2015lwa,Linch:2015fya} we analyzed the extension of worldsheet first-quantization of 3D string theory (S-theory), and its manifestly T-dual formulation (T-theory), to M-theory and F-theory on branes. 
Our approach \cite{Linch:2015fya} is unrelated to any other treatment of branes in that the brane coordinates $X(§)$ carry only indices that are simultaneously worldvolume and spacetime indices.  They are also selfdual differential forms (similar to \cite{Kutasov:1996vh,Kutasov:1996zm,Kutasov:1996fp}).  This results in section conditions involving not only spacetime (as for T-gravity \cite{Siegel:1993xq, Siegel:1993th, Siegel:1993bj} and F-gravity \cite{Berman:2011cg,Coimbra:2011ky, Berman:2012vc}) but also the worldvolume.  If extended to the full 10D type II string, this would allow for the first time an analysis of massive modes under the full STU-duality.

In this paper we describe the F-theory corresponding to type II strings in 4D from its formulation as a fundamental brane. 
This new theory is the complexification of the fundamental 5-brane theory \cite{Linch:2015fya} corresponding to the 3D type II string after reformulating the latter in bispinor notation. In addition to the target space sectioning constraint and Gau\ss{}'s law relating target and worldvolume coordinates, the Hamiltonian analysis reveals a constraint implying a new type of sectioning quadratic in derivatives on the {worldvolume}. 
Solving these conditions reduces the F-theory to M-, T- and S-theories, as required. 
Alternatively, we recover the original 5-brane theory by a double dimensional reduction (corresponding to wrapping the brane on a 6-torus and then compactifying).
By design, the theory's current algebra gives rise to C- and D-brackets that are covariant under the exceptional symmetry $E_{5(5)} =$ \spin55. After coupling to a general background, we verify explicitly that they reduce to the exceptional geometry brackets of F-gravity when truncated to massless modes. 

The remainder of this note is organized as follows: 
In section \ref{S:3D} we review the 5-brane theory corresponding to the 3D type II string in a formulation amenable to generalization to 4D. This is carried out in section \ref{S:4D} by complexifying the coordinates in the Lagrangian formulation. 
The three types of constraints relating target space and worldvolume coordinates to themselves and each other are derived. They and the currents form a closed subalgebra that is studied in section \ref{S:AB}. In section \ref{S:Sectioning} the constraints are solved, reducing F $\to$ M, T, and S. The double dimensional reduction recovering the 5-brane theory is also given. Our results are summarized in section \ref{S:Conclusions}.

%%%%%%%%%%%%%%%%%%%%%%%%%%%%%%%%%%%%%%%%%%%%%%%%%%%%%%%%%%%%%%%%
\section{3D revisited}
\label{S:3D}
Since the isotropy group \sp4C of the F-theory for the 4D superstring is the complexification of that for 3D, it will prove suggestive to review the 3D case here.  However, the spacetime coordinates of the 4D case are in the spinor representation of the isometry group $\mathrm O(5,5)$, which begs the use of spinor notation.  The result resembles the so-called ``K\"ahler-Dirac formalism" \cite{Iwanenko:1928aa, Lanczos:2005uh, Conway:1937zz, Kahler:1962} in that the gauge fields, gauge parameters, etc., are represented by bispinors (polyforms).

The Lagrangian for the selfdual 3-form on a 5-brane has manifest O(3,3) invariance \cite{Linch:2015fya}. In \spin33 $=$ \sl4R form, the 2-form gauge field is a real, traceless $4\times 4$ matrix (\spin33 2-form) with 1-form gauge transformation 
\begin{align}
\label{E:3Dgauge}
\delta Z_\alpha{}^\beta &= 
	\partial_{\alpha \gamma} \lambda^{\beta \gamma}
	- \partial^{\beta \gamma} \lambda_{\alpha \gamma} \cr
	&= 2 \partial_{\alpha \gamma} \lambda^{\beta \gamma}	
		- \tfrac12 \delta_\alpha^\beta \partial_{\gamma \delta} \lambda^{\gamma \delta}
~~~\mathrm{with}~~~
\lambda^T = - \lambda .
\end{align}
(Pairs of anti-symmetric spinor indices can be raised and lowered with $\tfrac12 \epsilon^{\alpha \beta \gamma \delta}$ and $\tfrac12 \epsilon_{\alpha \beta \gamma \delta}$.)
The fieldstrength has (anti-)selfdual parts which become symmetric bispinors
\begin{align}
F^{(+)}_{\alpha \beta} = \partial_{\gamma(\alpha}Z_{\beta)}{}^\gamma
~~~\mathrm{and}~~~
F^{(-)\alpha \beta} = \partial^{\gamma(\alpha}Z_\gamma{}^{\beta)}
\end{align}
satisfying the Bianchi identity
\begin{align}
\label{E:3DBianchi}
\partial^{\alpha \gamma} F^{(+)}_{\beta \gamma} 
	-  \partial_{\beta \gamma} F^{(-)\alpha \gamma} = 0.
\end{align}
This makes clear the infinite, repeating reducibility of $F\to Z \to \lambda \to \dots$ studied in detail in reference \cite{Berman:2012vc}.
So far, the structure is actually \gl4R covariant but invariance of the Lagrangian
\begin{align}
\label{E:3DLagrangian}
L= - \tfrac18\mathrm{tr} \, F^{(+)}F^{(-)} 
\end{align}
 reduces this to \sl4R. (The use of $\epsilon$ changes the GL(1) weight with implications for the associated (super)gravity; cf. \cite{Linch:2015lwa}.)

Reducing to the Hamiltonian formulation requires us to break \sl4R $\to$ \sp4R. Then
%\footnote{Although raising indices on $C_{\alpha \beta}$ introduces no sign ($C^{\alpha \gamma} C^{\beta \delta} C_{\gamma \delta} = C^{\alpha \beta}$), $\tfrac12 \epsilon^{\alpha \beta \gamma \delta} C_{\gamma \delta} =  - C^{\alpha \beta}$ (the singlet in ${\bf 6 \to 5\oplus 1}$ is anti-selfdual).} %end footnote
\begin{align}
\partial_{\alpha \beta} \to \partial_{\alpha \beta} + C_{\alpha \beta} \partial_\tau
~~~\mathrm{and}~~~
\partial^{\alpha \beta} \to \partial^{\alpha \beta} - C^{\alpha \beta} \partial_\tau ,
\end{align}
where now $C^{\alpha \beta} \partial_{\alpha \beta} = 0$. 
Upon lowering an index, the 2-form 
\begin{align}
Z_{\alpha \beta} = \tfrac12 (X_{\alpha \beta} + Y_{\alpha \beta})
~:~~X_{\alpha \beta} = X_{\beta \alpha}
~,~~Y_{\alpha \beta} = -Y_{\beta \alpha}
~,~~C^{\alpha \beta} Y_{\alpha \beta} = 0
\end{align}
decomposes into a symmetric part $X$ (2-form) and an antisymmetric, $C$-traceless part $Y$ (vector).
The fieldstrengths reduce to 
\begin{align}
F^{(+)}_{\alpha \beta} = 
	\dt X_{\alpha \beta}
	+\tfrac12 \partial_{\gamma(\alpha}X_{\beta)}{}^\gamma
	+\tfrac12 \partial_{\gamma(\alpha}Y_{\beta)}{}^\gamma
~~~\mathrm{and}~~~
F^{(-)}_{\alpha \beta} = 
	\dt X_{\alpha \beta}
	-\tfrac12 \partial_{\gamma(\alpha}X_{\beta)}{}^\gamma
	+\tfrac12 \partial_{\gamma(\alpha}Y_{\beta)}{}^\gamma ,
\end{align}
and the Lagrangian becomes
\begin{align}
L = - \tfrac14 (\dt X_{mn} + \partial_{[m} Y_{n]})^2 
	+\tfrac1{12} (\partial_{[p}X_{mn]})^2
.
\end{align}
The momentum conjugate to $Y$ is identically 0 whereas that conjugate to $X$ is 
\begin{align} 
P_{mn} = -{\delta S\over \delta \dt X{}^{mn}} = \dt X_{mn} + \partial_{[m}Y_{n]} .
\end{align}
The action in Hamiltonian form is expressed in manifestly SO(3,2)-covariant notation as
\begin{align}
S &= - \int \tfrac12 P_{mn} \dt X{}^{mn} \, d^5\sigma d\tau + \int H d\tau \cr
H&= \int \left[ \tfrac14 P_{mn}P^{mn} 
	+ \tfrac1{12}(\partial_{[p}X_{mn]})^2  
	+ Y^{m}\partial^nP_{mn}  \right]d^5\sigma .
\end{align}
The field $Y$ remains only as a Lagrange multiplier for the Gau\ss{} law constraint
\begin{align}
\U_m := \partial^nP_{mn} = 0.
\end{align}
With the Gau\ss{} law constraint taken into account, we can gauge away the Lagrange multiplier $Y\to 0$.

The stress-energy tensor in \sl4R notation is
\begin{align}
\T^{(+)}_{\alpha \beta \, \gamma\delta} = F^{(+)}_{\gamma [\alpha} F^{(+)}_{\beta] \delta} .
\end{align}
(This form implies the symmetries 
$\T_{\alpha \beta \, \gamma\delta} 
	= \T_{\gamma\delta\,\alpha \beta }  
	= - \T_{\beta \alpha  \, \gamma\delta}$
and the identity $\T_{\alpha [\beta \, \gamma\delta]} =0$.)
It decomposes into \sp4R $=$ \spin32 representations $\T_{mn}$, $\S^m$, and $\T$ ($\bf 21= 15 \oplus 5 \oplus 1$) of which only
\begin{align}
\S^r  = \tfrac18 \epsilon^{mnpqr} \dd_{mn} \dd_{pq} 
~~~\textrm{with current}~~~
\dd_{mn} := \tfrac14 \mathrm{tr} (\gamma_{mn} F^{(+)})
\end{align}
is manifestly \sl5R-covariant \cite{Linch:2015fya}.
Together, $\S$ and $\U$ form a closed subalgebra of the Virasoro algebra + Gau\ss{} law constraint with the larger-rank $E_{4(4)} =$ \sl5R symmetry. 
This is summarized in table \ref{T:3D}.

\begin{table}[h]
\begin{center}
{\renewcommand{\arraystretch}{1.3} %adds some padding
\begin{tabular}{|c|ccccc|}
\hline
&Lagrangian & $\longrightarrow$ & Hamiltonian & $\longrightarrow$ & Current Algebra\\
\hline
Symmetry&\spin33 = \sl4R 	& $\twoheadrightarrow$ &  \spin32 = \sp4R & $\hookrightarrow$ & \sl5R \\
Virasoro&$\T^{(+)}_{\alpha \beta \, \gamma\delta} $ && $\T_{mn}, \S^m, \T$ && $\S^m$ \\
Gau\ss{} &	&& $\U_m$ && $\U_m$ \\
\hline
\end{tabular}
} %end arraystretch
\end{center}
\begin{caption}{Symmetry breaking and enhancement in the 5-brane system (rank 4)}
\label{T:3D}
\footnotesize
The Lagrangian description of the dynamics preserves a larger symmetry than the Hamiltonian description. 
On the other hand, the Virasoro+Gau\ss{} algebra contains a kinematic subalgebra generated by $\S$ and $\U$ preserving a higher-rank exceptional symmetry.
\end{caption}
\end{table}

%%%%%%%%%%%%%%%%%%%%%%%%%%%%%%%%%%%%%%%%%%%%%%%%%%%%%%%%%%%%%%%%
\section{Worldvolume Action}
\label{S:4D}
We now give a covariant 4D theory by an appropriate complexification of the 3D case in spinor notation.  
The F-theory for the 4D string with global symmetry $E_{5(5)} =$ \spin55 is 16-dimensional with coordinates $X^\mu$ in the spinor representation. 
Since $X^µ$ reduces in \sp4C to a $\bf (4,Ð4)$, the 3D $Z_Œ{}^º$ of \gl4R must generalize (complexify) to $Z_Œ{}^{Àº}$ of \gl4C, now lacking both trace and reality conditions.

Using \gl4C, $\partial$ and $Z$ are both $4\times4$ complex matrices
\begin{align}
\partial_{\alpha \beta} 
~,~
\bar \partial_{\dt\alpha \dt\beta} = (\partial_{\alpha \beta} )*
~,~
Z_\alpha{}^{\dt \alpha}
~,~
\bar Z_{\dt \alpha}{}^{\alpha} = (Z_\alpha{}^{\dt \alpha})*
\end{align}
where the $\partial$'s are anti-symmetric ($\bf 6$ and $\overline{\bf 6}$), and $Z$ ($\bf 16_C$) is the complexification of $X$ (which is Hermitian $X^\dagger = X$). 
Then the gauge transformation generalizing (\ref{E:3Dgauge}) is
\begin{align}
\label{E:gaugeXf}
\delta Z_\alpha{}^{\dt \alpha} = 
	\partial_{\alpha \beta} \lambda^{(+)\beta \dt \alpha}
	-\bar \partial^{\dt\alpha \dt\beta} \lambda^{(-)}\hspace{-3mm}{}_{\alpha \dt \beta}
~~~\mathrm{with}~~~
\lambda^{(\pm)}{}^{\dagger} = - \lambda^{(\pm)}
\end{align}
anti-Hermitian gauge parameters ({\bf 16} and {\boldmath $16'$}):
$Z$ has the interpretation of a complex gauge 2-form with a complex 1-form gauge parameter.
From this, we form the Hermitian matrices
\begin{align}
\label{E:Fs}
F^{(+)} =  \bar \partial Z + \partial \bar Z
&â\Leftrightarrow~~~
~~F^{(+)}_{\dt \alpha \alpha}~ = \bar \partial_{\dt\beta \dt\alpha } Z_\alpha{}^{\dt \beta}
	+\partial_{\beta \alpha } \bar Z_{ \dt \alpha}{}^\beta \cr
F^{(-)} = \partial Z + \bar \partial \bar Z
&â\Leftrightarrow~~~
F^{(-)\alpha \dt \alpha} = \partial^{\beta \alpha } Z_\beta{}^{ \dt \alpha}
	+\bar \partial^{\dt\beta\dt\alpha } \bar Z_{\dt\beta}{}^{\alpha} .
\end{align}
These are invariant under the gauge transformation (\ref{E:gaugeXf}) provided
\begin{align}
\label{E:V0}
\V:= \tfrac i8 ( \partial_{\alpha \beta} \partial^{\alpha \beta} 
	- \bar \partial_{\dt \alpha \dt \beta} \bar \partial^{\dt \alpha \dt \beta} )= 0 .
\end{align}
This is our first section condition.
Assuming this, $F$ satisfies the Bianchi identity (cf. \ref{E:3DBianchi})
\begin{equation}
\partial^{\alpha \beta} F^{(+)}_{\dt \alpha \beta}
- \bar \partial_{\dt \alpha \dt \beta} F^{(-)\alpha \dt \beta} = 0  
\end{equation}
and its conjugate giving again the infinite, repeating reducibility 
${\bf 16 \oplus 16^\prime \to 16_C \to \dots}$ of $F\to Z \to \lambda \to \dots$ (cf. \cite{Berman:2012vc}).

As with the 5-brane, the Lagrangian 
\begin{align}
L = - \tfrac18 \mathrm{tr}\, F^{(+)}F^{(-)} 
\end{align}
reduces the symmetry \gl4C $\to$ \sl4C. 
We now reduce this further \sl4C $\to$ \sp4C so $\bf 6 = 5\oplus1$ and $\bf 16_C = 16 \oplus 16$ with $\overline{\bf 16} = {\bf 16}$. We define this reduction by
\begin{align}
\partial Z \to \partial Z + \tfrac12( \dt Z + i Z^\prime )
~~~\mathrm{and}~~~
\bar \partial Z \to \bar \partial Z + \tfrac12( \dt Z - iZ^\prime )
\end{align}
where now again $C^{\alpha \beta} \partial_{\alpha \beta}=0$ and similarly for the conjugate. 
Decomposing 
\begin{align}
Z=X+iY 
\end{align} 
for Hermitian $X$ and $Y$, we get the field strengths (free indices lowered)
\begin{align}
F^{(+)} &\to 
	[\dt X +i  (\partial - \bar \partial)Y]
	+ [Y^\prime - (\partial + \bar \partial)X]\cr
F^{(-)} &\to 
	[\dt X  + i (\partial - \bar \partial)Y]
	- [Y^\prime - (\partial + \bar \partial)X] 
.
\end{align}
The action reduces to
\begin{align}
S = - \tfrac 12 \int 
	\left\{
	\left[\dt X + \tfrac i2 (\partial - \bar \partial)Y\right]^2
	-\left[Y^\prime - \tfrac12 (\partial + \bar \partial)X\right]^2
	\right\} \, d^{12} \sigma
.
\end{align}
The momentum conjugate to $X$ becomes
\begin{align}
P_{\alpha \dt \alpha}:= - {\delta S \over \delta \dt X{}^{\alpha \dt \alpha}} 
	= \dt X_{\alpha \dt \alpha} 
		+ \tfrac i2 (\partial_{\alpha \beta}Y^\beta{}_{\dt \alpha}  - \bar \partial_{\dt \alpha \dt \beta} Y_\alpha{}^{\dt \beta}) .
\end{align}
Because of the form of the fieldstrengths, the action does not contain a $\dt Y{}^2$ term. Interpreting $X$ as the dynamical field, this means that in the Hamiltonian analysis of this system we should treat $\tau$ as the ``time'' parameter conjugate to the Hamiltonian. In this sense, $Y$ is not dynamical and we will gauge it to 0 presently.

The $\V$ constraint (\ref{E:V0}) reduces to
\begin{align}
\V \to \tfrac i8 (\partial_{\alpha \beta} \partial^{\alpha \beta} 
	- \bar \partial_{\dt \alpha \dt \beta} \bar \partial^{\dt \alpha \dt \beta})
	-\tfrac18 \partial_\tau \partial_\sigma
.
\end{align}
A partial solution of this constraint is obtained by setting
\begin{align}
\label{E:V1}
(\textrm{anything})^\prime = 0
~~~\textrm{and reducing}~~~
\V \to \tfrac i8 ( \partial_{\alpha \beta} \partial^{\alpha \beta} 
	- \bar \partial_{\dt \alpha \dt \beta} \bar \partial^{\dt \alpha \dt \beta} ) .
\end{align}
With this the action in Hamiltonian form becomes
\begin{align}
S &= - \int P_{\alpha \dt \alpha} \dt X{}^{\alpha \dt \alpha} \, d^{10}\sigma d\tau +  \int H \, d\tau  \\
H &=\tfrac12 \int \left[ 
	 P_{\alpha \dt \alpha} P^{\alpha \dt \alpha}
	+ (\partial_{\alpha \beta} X^\beta{}_{\dt \alpha} 
				- \partial_{\dt \alpha \dt \beta} X_\alpha{}^{\dt \beta} )^2 
	+ i Y_{\alpha}{}^{ \dt \alpha} (\partial^{\alpha \beta}P_{\beta \dt \alpha}  + \bar \partial_{\dt \alpha \dt \beta} P^{\alpha \dt \beta})
	\right] d^{10}\sigma,
\nonumber	
\end{align}
where we have normalized the volume of the gauge-fixed $\sigma$ direction to 1.
Note that this expression for the Hamiltonian cannot be rewritten with manifest \spin55 invariance (e.g. $P_{\alpha \dot \alpha}\to P_\mu$ is a chiral ten-dimensional spinor).
We interpret the field $Y$ as a Lagrange multiplier for the constraint 
\begin{align}
\label{E:U0}
\mathcal U_{\alpha}{}^{ \dt \alpha} = 
	\tfrac i2 (\partial_{\alpha \beta}P^{\beta \dt \alpha}  + \bar \partial^{\dt \alpha \dt \beta} P_{\alpha \dt \beta}) 
%\mathcal U^\mu := (\gamma_m)^{\mu \nu} \partial^m P_\nu.
\end{align}
generating a bosonic $\kappa$-symmetry; we use it to gauge $Y\to 0$.
After this is imposed, the fieldstrengths can be written in manifestly \spin55-covariant form
\begin{align}
\label{E:currents}
\dd_\mu := F^{(+)}_\mu = P_\mu + (\gamma_m)_{\mu\nu}\partial^m X^\nu 
~~~\mathrm{and}~~~
\tilde \dd_\mu := F^{(-)}_\mu  = P_\mu - (\gamma_m)_{\mu\nu}\partial^m X^\nu 
\end{align}
after combining \sl4C indices into the 16 $\times$ 16 Pauli matrices 
\begin{align}
(\gamma^m)^{\mu \nu}  = 
	\left( \begin{array}{cc}
			C^{\dt \alpha \dt \beta} (\gamma^m)^{\alpha \beta} & 0 \\
			0 & C^{\alpha \beta} (\gamma^m)^{\dt \alpha \dt \beta}
			\end{array} \right)
\end{align}
of \spin55.

The stress-energy tensor
\begin{align}
\T^{(+)}_{\alpha \beta \, \dt \alpha \dt \beta}  = F^{(+)}_{\dt \alpha [\alpha} F^{(+)}_{\dt \beta \beta]}
\end{align}
satisfies 
$\T^{(+)}_{\alpha \beta \, \dt \alpha \dt \beta} 
	= \T^{(+) \dagger}_{\alpha \beta \, \dt \alpha \dt \beta} 
	= - \T^{(+)}_{\beta \alpha  \, \dt \alpha \dt \beta}
	= - \T^{(+)}_{\alpha \beta \, \dt \beta \dt \alpha } $. 
It decomposes into \sp4C representations $\T_{mn}$, $\S^m$, and $\T$ 
(${\bf 36 = 25 \oplus10 \oplus 1}$)
with 
\begin{align}
\label{E:T}
\T_{\alpha \beta \, \dt \alpha \dt \beta}  = F^{(+)}_{\dt \alpha [\alpha} F^{(+)}_{\dt \beta \beta]} 
	- C\textrm{-traces}
~,~~
\S^m = \tfrac14 \dd \gamma^m \dd 
~,~~
\T = \tfrac14 C^{\alpha \beta} C^{\dt \alpha \dt \beta} F^{(+)}_{\dt \alpha \alpha} F^{(+)}_{\dt \beta \beta} .
\end{align}
Again only the $\S$ current can be written in manifestly \spin55-covariant form:
The subalgebra of currents $\S$, $\U$, $\V$ is manifestly \spin55 covariant even thought the Hamiltonian description of the dynamics preserves only the \sp4C subgroup. 
We summarize this in table \ref{T:4D} (cf. table \ref{T:3D}). 

\begin{table}[htb]
\begin{center}
{\renewcommand{\arraystretch}{1.3} %adds some padding
\begin{tabular}{|c|ccccc|}
\hline
&Lagrangian & $\longrightarrow$ & Hamiltonian & $\longrightarrow$ & Current Algebra\\
\hline
Symmetry&Spin(6;{\bf C}) = \sl4C 	& $\twoheadrightarrow$ & ${\rm Spin}(5;{\bf C})$ = \sp4C & $\hookrightarrow$ & Spin(5,5) \\
Virasoro&$\T^{(+)}_{\alpha \beta \, \dt \alpha \dt \beta}$ && $\T_{mn}, \S^m, \T$ && $\S^m$ \\
Gau\ss{} &	&& $\U_{\alpha \dt \alpha}$ && $\U^\mu$ \\
Laplace &	 $\V$   && $\V, \partial_\sigma$ && $\V$ \\
\hline
\end{tabular}
} % end arraystretch
\end{center}
\begin{caption}{Symmetry breaking and enhancement in the 4D system (rank 5)}
\label{T:4D}
\footnotesize
The Lagrangian description of the dynamics preserves a larger symmetry than the Hamiltonian description but again there is a kinematic subalgebra of the Virasoro+Gau\ss+Laplace algebra preserving a higher-rank symmetry. 
Note that in this case the Lagrangian group Spin(6;{\bf C}) is not a subgroup of the full symmetry group \spin55. 
\end{caption}
\end{table}

%%%%%%%%%%%%%%%%%%%%%%%%%%%%%%%%%%%%%%%%%%%%%%%%%%%%%%%%%%%%%%%%
\section{Algebras and Backgrounds}
\label{S:AB}
We now give an independent way to derive $\S \to \U\to \V$ that could be useful in cases in which we do not know the covariant action. 
(This method is simpler than finding $\U$ and $\V$ by closing the $\S$ current algebra.)
The constraint $\S$ (\ref{E:T}) is defined in terms of $\dd$. Defining the analogous $\tilde \S$ in terms of $\tilde \dd$, 
\begin{align}
\label{E:StoU}
\S^m - \tilde \S^m = -i (\partial^m X^\mu) P_\mu + \mathcal O (\U)
\end{align}
(cf.\ \cite{Linch:2015fya}) we find the $\U$ constraint (\ref{E:U0}) in the form $(\sla \partial X)^\mu P_\mu$.
Similarly,
\begin{align}
\U = \tfrac12 \sla \partial (\dd + \tilde \dd) 
~~~\mathrm{and}~~~
\sla \partial (\dd - \tilde \dd) = \mathcal O (\V)
\end{align}
we find $\V$ (\ref{E:V1}) appearing as $\partial^m X^\mu \partial_m$. 
Just as $\S$ generates worldvolume coordinate transformations, $\U$ generates residual gauge transformations. (Both generate what is left of local invariances once $\partial_\tau$ is thrown away.)

We next examine the current algebras.  The covariant derivatives and symmetry currents (\ref{E:currents})
$$
%\begin{align}
\dd_\mu 
	=P_\mu + (\gamma_m)_{\mu\nu} \partial^m X^\nu
	%+ i (\theta \gamma_{mnp} \theta)(\gamma^{mn}\partial^p\theta)_\mu 
¼,ââ
\tilde \dd_\mu 
	=P_\mu - (\gamma_m)_{\mu\nu} \partial^m X^\nu
%\end{align}
$$
are bosonic, despite their resemblance to supersymmetry currents. 
Using the Poisson bracket 
\begin{align}
\left[ P_\mu(1) , X^\nu(2) \right] = - i \delta_\mu^\nu \delta(1 - 2)
\end{align}
for the momentum $P_\mu$ conjugate to $X^\mu$, the brackets of the currents are 
\begin{align}
\label{E:currentAlgebra}
[ \dd_\mu(1) , \dd_\nu(2) ] & = 2i (\gamma_m)_{\mu \nu} \partial^m \delta(1-2) \cr
[ \dd_\mu(1) , \tilde \dd_\nu(2) ] &= 0 \cr
[\tilde \dd_\mu(1) ,\tilde \dd_\nu(2) ] &= -2i (\gamma_m)_{\mu \nu} \partial^m \delta(1-2)  .
\end{align}

The pure-spinor-like constraint (\ref{E:T}) \cite{Linch:2015fya}
\begin{align}
\label{E:S}
\mathcal S^m = \tfrac14 (\gamma^m)^{\mu\nu} \dd_\mu \dd_\nu 
\end{align}
has Poisson bracket with the current given by
\begin{align}
[ \mathcal S^m(1), \dd_\mu(2) ] &= i ( \gamma_n \gamma^m)_\mu{}^\nu \partial^n \delta(1-2) \dd_\nu(1).
\end{align}
Using 
\begin{align}
\partial^p \delta(1-2) A(1)B(2) = 
	\partial^p \delta(1-2) AB\tfrac12((1)-(2)) 
	+ \tfrac 12 A \stackrel{\leftrightarrow}{\partial^{\,p}} B ,
\end{align}
we find the algebra
\begin{align}
[ \mathcal S^m, \mathcal S^n] &= 
	2 i \partial^{(m} \delta%(1-2) 
		\mathcal S^{n)}%\tfrac12((1)+(2)) 
	- 2 i \eta^{mn}  \partial^p \delta(1-2) \mathcal S_p%\tfrac12((1)+(2)) 
	 -\tfrac i2 \delta%(1-2) 
	 	\left[ 2\partial^{[m} \mathcal S^{n]}
			+ (\dd \gamma^{mn} \mathcal U) \right] 
\end{align}
similar to that of \cite{Linch:2015fya}.
Here the $\partial \delta$ terms are evaluated at $\tfrac12((1)+(2))$, $\dd \gamma^{mn} \mathcal U = \dd_\nu (\gamma^{mn})^\nu{}_\mu  \mathcal U^\mu$ with the bosonic $\kappa$-symmetry generator 
\begin{align}
\label{E:U}
\mathcal U^\mu := (\gamma_m)^{\mu \nu} \partial^m \dd_\nu ,
\end{align}
found previously in (\ref{E:U0}).
The existence of $\U$ immediately implies another constraint: $(\gamma_m)_{\mu \nu} \partial^m \U^\nu = \V \dd_\mu$ where 
\begin{align}
\label{E:V}
\V := \eta_{mn} \partial^m \partial^n .
\end{align}
Thus, we recover the condition (\ref{E:V0}) required by gauge invariance of the Lagrangian description.

The algebra of constraints generated by $\S$, $\U$, and $\V$ closes, and the new constraint gives rise to a third section condition, this time on the {\em worldvolume}. 
This new constraint implies the gauge invariance 
\begin{align}
\label{E:gXf}
\delta X^\mu = (\gamma_m)^{\mu \nu} \partial^m \lambda_\nu .
\end{align}
and the gauge-for-gauge transformation
\begin{align}
%\label{E:gXf}
\delta \lambda_\mu = (\gamma_m)_{\mu \nu} \partial^m \lambda^\nu .
\end{align}
Clearly, the gauge invariance is infinitely reducible. 

The worldvolume derivative of a function $f=f(X)$ is given by
\begin{align}
\label{E:partials}
\partial^m f &= \partial^m X^\mu \partial_\mu f = \tfrac12 (\gamma^m \gamma_n + \gamma_n \gamma^m)^\mu{}_\nu \partial^n X^\nu \partial_\mu f 
\cr&
\equiv \tfrac12 (\gamma^m \gamma_n)^\mu{}_\nu \partial^n X^\nu \partial_\mu f 
~~(\mathrm{mod} ~\U)~
%\cr&
	= \tfrac14 (\gamma^m)^{\mu \nu} (\dd_\nu -\tilde \dd_\nu)  \partial_\mu f \cr
	&\equiv \tfrac14 (\gamma^m)^{\mu \nu} \partial_\mu f \dd_\nu
~~(\mathrm{mod} ~\tilde \dd) .
\end{align}
in agreement with (\ref{E:StoU}). Using this, we derive the Poisson bracket of two vector fields $V_i = V^\mu_i \dd_\mu$ for ${}_i={}_{1,2}$. Modulo second class constraints and sectioning this gives the C-bracket \cite{Siegel:1993th} 
(again with the $\partial \delta$ term evaluated at $\tfrac12((1)+(2))$)
\begin{align}
\label{E:[VV]}
[V_1(1), V_2(2)] &= 
	2i \partial^m \delta  V_1\gamma_m V_2
	- i\delta
		\left[  \delta_\rho^\mu \delta_\sigma^\nu 
			-\tfrac14 (\gamma_m)_{\rho \sigma} (\gamma^m)^{\mu \nu} \right]
			V_{[1}^\rho \partial_\mu V_{2]}^\sigma \dd_\nu .
\end{align} 
The truncation of this bracket to massless modes reproduces the ``exceptional Courant bracket'' of reference \cite{Coimbra:2011ky,Berman:2012vc}. %cf. eqs. (2.10) and (2.18) of Berman (and 2.28 of Coimbra)

We now include backgrounds by dressing the covariant derivative 
\begin{align}
\dd_\alpha = e_\alpha{}^\mu \dd_\mu.
\end{align}
Using (\ref{E:[VV]}), we find that under worldsheet reparameterizations $\delta_\lambda \dd_\alpha = [i\int \lambda^\mu\dd_\mu, \dd_\alpha]$, the vielbein transforms according to 
\begin{align}
\delta_\lambda e_\alpha{}^\mu &= 
	\lambda^\nu \partial_\nu e_\alpha{}^\mu
	-  e_\alpha{}^\nu \partial_\nu \lambda^\mu
	+ \tfrac14 (\gamma_m)_{\rho \sigma} (\gamma^m)^{\mu \nu} 
		e_{\alpha}{}^\rho \partial_\nu \lambda^\sigma
\end{align}
in agreement with the results of \cite{Coimbra:2011ky, Berman:2012vc}.
The commutation relations in a general background are 
\begin{align}
\label{E:generalBackground}
[\dd_\alpha(1),\dd_\beta(2)] = 2i \partial^m \delta(1-2) g_{\alpha \beta m}\tfrac12((1)+(2)) -i\delta(1-2)f_{\alpha \beta}{}^\gamma\dd_\gamma
\end{align}
where 
\begin{align}
\label{E:gf}
g_{\alpha \beta m} := e_\alpha{}^\mu  (\gamma_m)_{\mu\nu} e_\beta{}^\nu
~~~\mathrm{and}~~~
f_{\alpha \beta}{}^\gamma := c_{[\alpha \beta]}{}^\gamma +\tfrac12 c_{\delta [\alpha}{}^\varepsilon (g^{\gamma \delta m}g_{\beta] \varepsilon m}) .
\end{align}
Here the $g$'s are defined by the first equation and the generalization 
\begin{align}
\partial^m f = -\tfrac14 g^{\alpha \beta m} \partial_\alpha f \dd_\beta
\end{align}
of (\ref{E:partials}), and the ``anholonomy''-type coefficients (not anti-symmetric) are defined by
\begin{align}
c_{\alpha \beta}{}^\gamma:=(e_\alpha e_\beta{}^\mu)e_\mu{}^\gamma .
\end{align}
The Bianchi identity $[[\dd_{(\alpha},\dd_\beta], \dd_{\gamma)}] = 0$ then implies the relations
\begin{align}
f_{\gamma(\alpha}{}^\delta g_{\beta)\delta m} &= 2e_\gamma g_{\alpha \beta m} - e_{(\alpha} g_{\beta)\gamma m}\cr
\tfrac16 e_{[\alpha}f_{\beta\gamma}{}^\varepsilon g_{\delta ]\varepsilon m} &= \tfrac18 f_{[\alpha \beta}{}^\varepsilon  f_{\gamma \delta]}{}^\varphi  g_{\varepsilon \varphi m} .
\end{align}
These results should be compared with the analogous expressions in reference \cite{Siegel:1993th}.

%%%%%%%%%%%%%%%%%%%%%%%%%%%%%%%%%%%%%%%%%%%%%%%%%%%%%%%%%%%%%%%%
\section{Sectioning}
\label{S:Sectioning}
New section conditions are obtained by replacing string coordinates with 0-modes \cite{Kugo:1992md}. In addition to the new section condition from Gau\ss{}'s law found in reference \cite{Linch:2015fya}, there is yet another type of section condition on the worldvolume coming from the Laplace constraint (\ref{E:V}). We collect these conditions in the following table:
\begin{subequations}
\label{subsections}
\setlength{\fboxsep}{10pt}
\begin{empheq}[box=\fcolorbox{green}{white}]{align}
&~\hbox{Virasoro} &\mathcal S^m =¼& \tfrac14 (\dd\gamma^m\dd) && \\
\label{E:dimRed}
&\begin{array}{l} \hbox{dimensional} \\ \hbox{reduction}\end{array} & \on\circ\S{}^m :=¼& (p \gamma^m P) & \mathcal U^\mu =¼& (\sla\partial P)^{\mu} &\\
\label{E:section}
& \begin{array}{l} \hbox{section} \\ \hbox{condition}\end{array} & \under\circ\S{}^m :=¼& \tfrac12 (p \gamma^mp ) & \under\circ\U{}^\mu :=¼& (\sla\partial p)^\mu& 
\V :=¼&\partial^m\partial_m
%\end{align}
\end{empheq}
\end{subequations}

Since we now have 3 different types of section conditions (``strong constraints"), this might be a good place to review the method of their solution. The basic point is that these conditions are applied at 2 independent points in ``function space":  They take the generic form
\begin{equation}
»»A = 0â\hbox{and}â(»A)(»B) = 0
\end{equation}
for arbitrary functions $A$ and $B$ and with various reductions (contractions, symmetrizations, etc.) on the indices.  Thus in momentum space
\begin{equation}
pp' = 0
\end{equation}
where $p$ and $p'$ may or may not be at the same point in function space. (In fact, our function space is disjoint, having momenta for both the worldvolume and spacetime:  In particular, for the $\under\circ\U$ section condition one of the momenta is in the worldvolume and the other in spacetime \cite{Linch:2015fya}.)

So we have not only a quadratic constraint $pp=0$, but also a bilinear one $pp'=0$.  For example, for T-theory we have the universal constraint $pÉp'=0$, taking the inner product with the signature of the T-symmetry group O(D,D).  For the quadratic constraint the most general solution is to pick a lightlike basis where the O(D,D) metric is block off-diagonal, then choose a frame where $p$ has vanishing components in one of the 2 subspaces (``section") corresponding to this block decomposition.  The bilinear constraint is then solved by taking $p$ in such a frame and finding that $p'$ must be not only of the same form but in the same frame (i.e., in the same subspace).  Conversely, given this $p'$ we find that we could have chosen another $p$, but still in this same subspace.  Thus although the frame is arbitrary, it is the same over all function space:  All fields live on the same D-dimensional subspace of the original 2D-dimensional space. (This reduces T-theory to S-theory.)

Another example is the $\under\circ\U$ constraint $p^n p'_{mn}=0$ considered previously for the F-theory of the 3D string.  It is only bilinear, since $p^m$ is in the worldvolume while $p'_{mn}$ is in spacetime.  Because this constraint (and the whole formulation) is GL(5) covariant, we can always choose a frame where $p^m$ is in one particular direction, even before considering constraints.  This directly kills all of $p'_{mn}$ carrying that index.  Conversely, this general solution for $p'_{mn}$ implies that $p^m$ can only point in that one direction, not only for that function, but for any function on the worldvolume.  Thus again the frame is arbitrary, but applies to all functions of either the worldvolume or spacetime.  (This reduces F-theory to T-theory.)

We will now carry out this reduction from F to M, T, and S for the 4D type II string. (See \cite{Linch:2015fya} for the corresponding analysis of the 3D type II string.) The solution is represented schematically in the F-theory diamond of figure \ref{F:STMF}. 

\begin{figure}[h]
\begin{align}
\xymatrix{
	& \fbox{ \txt{ {\bf F}($X$) \\ $E_{n(n)}/H_n$} \ar[dl]_{\S} \ar[dr]^{\U\, \&\, \V} }&\cr
\fbox{ \txt{ {\bf M}($X$)\\ $\rm GL(D+1)/O(D,1)$} \ar[dr] }&
		& \fbox{ \txt{ {\bf T}($X$)\\ ${\rm O(D,D)/[O(D-1,1)]}{}^2$} \ar[dl] } \cr
	&\fbox{ \txt{ {\bf S}($X$) \\ $\rm GL(D)/O(D-1,1)$}}&
}
\nonumber
\end{align}
\begin{caption}{F-, M-, and T-theories associated to type II string (S-theory) on $X$.}
\label{F:STMF}
\footnotesize
When the dimension of $X$ is D $=$ 3 or 4, there is associated to the D-dimensional type II supergravity $S(X)$ on $X$ a $(\mathrm D+1)$-dimensional N = 1 supergravity theory $M(X)$ and a D-dimensional, manifestly T-duality invariant supergravity $T(X)$. 
Each of these results from a theory $F(X)$ by solving the $\S$ constraint or $\U$ and $\V$ constraints, respectively \cite{Linch:2015lwa}. 
\end{caption}
\end{figure}

%%%%%%%%%%%%%%%%%%%%%%%%%%%%%%%%
\subsection{Subsectioning F\texorpdfstring{$\to$}{\textrightarrow}M}

We now solve the reduction conditions (\ref{E:dimRed}) and apply the logic above to the section constraints (\ref{E:section}).
We start with $\S$ conditions corresponding to the reduction F $\to$ M.

To solve the reduction and section conditions, we break \spin55 $\to$ \gl5R $=$ \sl5R $\times$ \gl1R. This is the same as the usual argument for O(2$n$) $\to$ U($n$) (but with split signature and real representations) so we suppress the details.
Decomposing
%the Pauli matrices become 
%\begin{align}
%(\gamma^P)^{\mu\nu} \to 
%	\left( \begin{array}{ccc}
%%
%	(\gamma^P)^{++} & (\gamma^P)^+{}_n & 	(\gamma^P)^{+ nn^\prime}  \\
%	(\gamma^P)_m{}^+ & (\gamma^P)_{mn} & (\gamma^P)_m{}^{nn^\prime}  \\	
%	(\gamma^P)^{mm^\prime +} & (\gamma^P)^{mm^\prime}_{n} & (\gamma^P)^{mm^\prime nn^\prime}	
%%	
%	\end{array}\right)
%=
%	\left( \begin{array}{ccc}
%%
%	0 & \delta^P_n & 	0  \\
%	\delta^P_m & 0 & \delta_m^{P nn^\prime}  \\	
%	0 & \delta^{Pmm^\prime}_n & \epsilon^{P mm^\prime nn^\prime}	
%%	
%	\end{array}\right)
%\end{align}
$P_\mu \to P^+, P^m, P_{mn}$ (${\bf 16 = 1\oplus 5 \oplus 10^\prime}$), and similarly for 0-modes, $\on\circ\S$ and $\under\circ\S$ become
\begin{align}
\label{E:FtoM}
\on\circ\S{}^m = (p\gamma^m P) 
~~~&\longrightarrow~~~
\begin{cases}
\on\circ\S{}^r =  p^+ P^r + p^rP^+ + \tfrac12 \epsilon^{mnpqr} p_{mn} P_{pq} \\
\on\circ\S{}_m = p^nP_{mn}  
\end{cases}
\cr
\under\circ\S{}^m = (p\gamma^m p) 
~~~&\longrightarrow~~~
\begin{cases}
\under\circ\S{}^r =  p^+ p^r + \tfrac14 \epsilon^{mnpqr} p_{mn} p_{pq} \\
\under\circ\S{}_m = p^np_{mn}  
\end{cases}.
\end{align}
(At this point, and for the remainder of this section only, the $m,n,\dots$ indices have become {\bf 5}'s.)
First applying the section conditions bilinearly, we find the solution
\begin{equation}
p^+ = p^{mn} = 0
\end{equation}
leaving only $p^m$.  (Other maximal solutions correspond to a different frame for breaking to GL(5).)
We then find similarly for the reduction conditions
\begin{equation}
P^+ = P^{mn} = 0.
\end{equation}

%%%%%%%%%%%%%%%%%%%%%%%%%%%%%%%%
\subsection{Subsectioning F\texorpdfstring{$\to$}{\textrightarrow}T}
Solving the $\U$ and $\V$ constraints reduces F $\to$ T. 
Unlike the 3D case reviewed in section \ref{S:3D}, the existence of the $\V$ condition (in combination with $\U$) now restricts the one direction of the $§$ derivative $»$ to be lightlike:
\begin{equation}
\V = »^2 = 0âÜâ» = »^+
\end{equation}
(The symmetry for this theory was only SO and not GL.)
The $\U$ and $\under\circ\U$ constraints thus reduce to
\begin{equation}
\under\circ\U = -»^+©^-p = 0âÜâ©^-p = 0
\end{equation}
\begin{equation}
\U = -»^+©^-P = 0âÜâ©^-P = 0
\end{equation}
So we are left with a single $§$ (in addition to $ $) and half (8) of the $X$'s, i.e., a string with twice (of 4) the dimensions (T-theory).

$\S$ now reduces to the usual for T-theory; solving also these constraints therefore gives the 4D string on the worldsheet.

%%%%%%%%%%%%%%%%%%%%%%%%%%%%%%%%
\subsection{Double Dimensional Reduction 4D \texorpdfstring{$\to$}{\textrightarrow} 3D}
Instead of solving constraints, we can perform the double dimensional reduction
\begin{align}
\label{E:wrapping}
P^+ \to 0
~,~~
P^m\to 0 
~,~~
\partial_m \to 0
\end{align}
(and similarly for their 0-modes) directly on the decomposition (\ref{E:FtoM}).
Then it is easy to see that what remains of the constraints is precisely the reduction and section conditions of the F-theory 5-brane for the 3D string \cite{Linch:2015fya}.
%Note that with $p^+=0$, the $\U_m$ condition becomes the Gau\ss{} law for the fundamental 5-brane.
In other words, the F-theory for the 4D type II string contains a subalgebra of constraints defined by the worldvolume $\bf 5$ and the spacetime ${\bf 10^\prime}$ that generates the F-theory algebra for the 3D type II string. Of course it is true that the 3D type II string is embedded in the 4D type II string but this observation implies that the entire rank 4 F-theory diamond (fig. \ref{F:STMF}) embeds into that of rank 5.

%%%%%%%%%%%%%%%%%%%%%%%%%%%%%%%%%%%%%%%%%%%%%%%%%%%%%%%%%%%%%%%%
\section{Conclusions}
\label{S:Conclusions}
In this paper we described the fundamental theory giving rise to the F-theory embedding the four-dimensional type II string 
(corresponding to the split form of the rank 5 exceptional group $E_{5(5)}=\mathrm{Spin}(5,5)$) 
as a complexification of that of the fundamental 5-brane of the 3D string \cite{Linch:2015fya}. 
This description requires three types of section condition: In addition to the original section condition ($\S$) on the target space \cite{Coimbra:2011ky, Berman:2012vc} and another  ($\U$) relating target space to worldvolume \cite{Linch:2015fya}, there is now also a third section condition  ($\V$) on the worldvolume itself. 
The analysis of these constraints shows that the 3- and 4-dimensional theories are related by double dimensional reduction (\ref{E:wrapping}).

The structure of these theories is such that the Lagrangian theory is invariant under an {\it a priori} unknown symmetry group $L_n$ that is broken to the subgroup $H_n$ in the Hamiltonian description. This subgroup is also the (split form of the) maximal compact subgroup of the split form $E_{n(n)}$. We represent this in table \ref{T:LHE}.
The algebra (\ref{E:currentAlgebra}) of the currents is manifestly $E_{n(n)}$-covariant as is the ``kinetic'' subalgebra of the full Virasoro+Gau\ss{}+Laplace algebra of constraints that is generated by $\S$, $\U$, and $\V$ (eqs. (\ref{E:S}), (\ref{E:U}), and (\ref{E:V}), resp. and cf. table \ref{T:4D}).

\begin{table}[h]
\begin{center}
{\renewcommand{\arraystretch}{1.2} %adds some padding
\begin{tabular}{|c|ccccccc|}
\hline
D 	& Lagrangian $L_n$	& $\to$ & Hamiltonian $H_{n}$ & $\to$ & Currents $E_{n(n)}$ &$\sigma$ & $X$ \\
\hline
3 & \spin33 = \sl4R 	&& \spin32 = \sp4R 	&& \sl5R & ${\bf 5^\prime}$ & ${\bf 10}$  \\  
4 	& ${\rm Spin}(6; {\bf C})$ = \sl4C && ${\rm Spin}(5; {\bf C})$ = \sp4C && \spin55 &${\bf 10}$& ${\bf 16}$ \\  
\hline
\end{tabular}
} %end arraystretch
\end{center}
\begin{footnotesize}
\begin{caption}{Symmetry groups of fundamental F-theory branes}
\label{T:LHE}
\footnotesize
The symmetry manifested by the Lagrangian and Hamiltonian formulations of the fundamental branes of F-theory corresponding to type II strings in $\mathrm D=$ 3 and 4 dimensions. The rank of the global exceptional symmetry $n=\mathrm D +1$ and the $E_{n(n)}$ representations of the worldsheet ($\sigma$) and target ($X$) coordinates are given in the last two columns.
\end{caption}
\end{footnotesize}
\end{table}

Clearly, it is of interest to extend this analysis to higher rank. The next classical superstring in the series is the 6D type II string corresponding to the maximal global symmetry $E_{7(7)}$. In this case the na\"ive dimension of the worldvolume exceeds that of the target so we expect the new worldvolume section condition (and possibly new constraints) to play a role in cutting this dimension down. 
Since these cases correspond to superstrings, supersymmetrization of our brane systems may give insight into the fundamental theories corresponding to these higher-dimensional F-theories. 

%%%%%%%%%%%%%%%%%%%%%%%%%%%%%%%%%%%%%%%%%%%%%%%%%%%%%%%%%%%%%%%%
\section*{Acknowledgements}
W{\sc dl}3 is partially supported by the U{\sc mcp} Center for String \& Particle Theory and National Science Foundation grants PHY-0652983, and PHY-0354401. % WDL3
W{\sc s} is supported in part by National Science Foundation grant PHY-1316617. % WS

%%%%%%%%%%%%%%%%%%%%%%%%%%%%%%%%%%%%%%%%%%%%%%%%%%%%%%%%%%%%%%%%
%%%%%%%%%%%%%%%%%%%%%%%%%%%%%%%%%%%%%%%%%%%%%%%%%%%%%%%%%%%%%%%%
%\newpage

%\small
%\footnotesize
\baselineskip=15pt
\bibliography{/Users/wdlinch3/Dropbox/Rashoumon/LaTeX/BibTex/BibTex}
\bibliographystyle{unsrt}

\end{document}